# IoT Devices Proximity Authentication In Ad Hoc Network Environment

Ali Abdullah S. AlQahtani
*Computer Systems Technology*
*North Carolina A&T State University*
Greensboro, North Carolina
alqahtani.aasa@gmail.com

Hosam Alamleh
*Computer Science*
*University of North Carolina Wilmington*
Wilmington, North Carolina
hosam.amleh@gmail.com

Baker Al Smadi
*Computer Science*
*Grambling State Univeresity*
Grambling, Louisiana
bakir_smadi@hotmail.com

*Abstract*—Internet of Things (IoT) is a distributed communication technology system that offers the possibility for physical devices (e.g., vehicles, home appliances sensors, actuators, etc.), known as *Things*, to connect and exchange data, more importantly, without human interaction. Since IoT plays a significant role in our daily lives, we must secure the IoT environment to work effectively. Among the various security requirements, authentication to the IoT devices is essential as it is the first step in preventing any negative impact of possible attackers. Using the current IEEE 802.11 infrastructure, this paper implements an IoT devices authentication scheme based on something that is *in the IoT device's environment* (i.e., ambient access points). Data from the broadcast messages (i.e., beacon frame characteristics) are utilized to implement the authentication factor that confirms proximity between two devices in an ad hoc IoT network.

## I. INTRODUCTION

THE Internet over the last four decades has developed from peer-to-peer networking (P2P), world-wide-web (WWW), and mobile-Internet to the IoT. The IoT is a network that might consist of animals, people, objects, physical devices (e.g., home appliance sensors, actuators, vehicles, digital machines, etc.) that can collect and exchange data with each other without human intervention. The way of communication in an IoT network can be between people, between people and devices, and between devices themselves, also called machine-to-machine (M2M).

The IoT environment has three major components; IoT devices, Cloud, and Client applications. IoT devices are responsible for sensing and measuring the world around them, taking local action as necessary (turning off/on or opening/closing an object, sharing data, etc.). The cloud is where the powerful applications reside and can collect data from IoT devices, combine IoT device data with other data sources, perform data analytics to reveal trends, identify problems, and predict the future. The client applications enable users to access and view data processed in the cloud issue commands to remote IoT devices.

In 2015, approximately 15 billion IoT devices worldwide were in use [1], which was doubled in 2020 (approximately around 31 billion) and might be around 60 billion by 2024 [2].

In general, IoT is continuously evolving and has become an actual attractive area of attack for hackers. The number of cyber-attacks on IoT devices is raised by 600% in only one year, from 2016 to 2017, corresponding to 6000 and 50,000 reported attacks, respectively [3]. IoT devices are usually subject to Distributed Denial-of-Service (DDoS) and ransomware attacks due to the fact that these attacks take advantage of storage limitations and their internet-supported connectivity [4] which grants hackers the opportunity to interact with devices remotely.

One way IoT networks can mitigate cyber-attacks is to establish authentication, which is based on trusting the other end before communicating with. As of today, there is a number of ways to establish authentication in IoT networks:

- **On-way authentication:** in the case before two IoT devices start to communicate with each other, only IoT device authenticates itself to the other, while the other IoT device will not be authenticated.
- **Two-way authentication:** both IoT devices must authenticate themselves to each other prior to the communication.
- **Three-way authentication:** a service provider is involved in this type, which authenticates the two IoT devices and assists them to authenticate each other.
- **Distributed:** this method utilizes a distributed authentication technique between the IoT devices prior to the communication.
- **Centralized:** a trusted third end is utilized to distribute and manage the authentication certificates

used.

IoT devices can be connected to a centralized network, in which all devices are connected directly the gateway. Alternatively, IoT devices can be connected to each other in ad hoc networks, where communications between IoT devices is used to relay information of other devices to the gateway.

We proposes a technique to authenticate IoT devices in ad hoc networks to verify proximity. This is done in a way that only devices within a certain distance from other authenticated IoT devices will be able to connect to the network. Meanwhile, devices that are far from an authenticated device or not physically in the area will fail in the proximity authentication. The proposed system enforces the security in ad hoc IoT networks.

## II. CONTRIBUTION

Using the current IEEE 802.11 infrastructure, this paper implements an IoT devices proximity authentication scheme in IoT ad hoc networks. This based on something that is *in the IoT device's environment* (i.e., ambient access points). In this paper, data (a Wi-Fi *footprint*)from the broadcast messages are utilized to implement the proximity by determining whether two devices are within a certain range for an authenticated IoT device unobtrusively.

## III. GUIDE TO PAPER

This paper is organized as follows; *Section IV* reviews the previous research on IoT devices authentication. *Section V* describes the proposed scheme system for beacon frame-based IoT devices authentication. The experiment of the proposed system and results are presented in *Section VI*. The proposed scheme is analysed in *Section VII*. Security analysis is discussed in *Section VIII*. Lastly, *Section IX* illustrates the conclusion for the proposed method.

## IV. RELATED WORK

This section provides an analytical overview of the literature proposing proximity-based authentication. Some systems use GPS for proximity verification [5], [6]. This is mainly done by comparing the GPS location calculated at two device. Few researches propose using GPS for location verification. However, using GPS suffers from performance limitations indoors [7]. Other systems use wireless sensor network for localization [8]. Moreover, utilizing wireless sensor network requires additional hardware installed on IoT devices, which can be costly to deploy and maintain. On the other hand, Bluetooth had been a popular choice for proximity-based authentication [9], [10]. In addition, the weakness in these approaches is that Bluetooth typically has a short-range and requires additional hardware that is not always guaranteed to be in the infrastructure or in every user's device [7]. Wi-Fi is a popular solution for proximity-based authentication are ubiquitous and widely used for connectivity in many users [11]–[13] and IoT devices. A few works use channel characteristics to verify proximity using signal information received from access points such as SSID, MAC address, and Received Signal Strength Indicator (RSSI) [14], multipath profiles [15], or mathematical modeling on channel characteristics [10], [16]. The proposed system uses Wi-Fi measurements to achieve authentication inthe context of ad hoc IoT networks. This can be doneby utilizing the role of Representational State Transfer (REST) API in the IoT Systems, which is able to record and count everything [17]. The proposed system provides a complete framework for IoT devices proximity verification and management, which is done by utilizing Wi-Fi Beacon frame information to perform proximity-based authentication.

## V. THE PROPOSED SCHEME

In this section, we will present the proposed scheme in detail. The proposed scheme aims to authenticate new IoT devices joining and ad hoc network. The type of authentication that is carried out is proximity-based. This aims to prevent adding new IoT nodes that are outside the "allowed area". This would also eliminate addingfake IoT nodes, or adding IoT node outside the allowed physical boundaries. The proposed system utilizes Wi-Fi beacons in the area to achieve this goal. In general, Wi-Fi access points periodically broadcast their own beacon frame, including Service Set Identifier (SSID) and Basic Service Set Identifier (BSSID). Moreover, Using the Wi-Fi footprint, we can measure the RSSI value of every presented access point in the location.

Before a new IoT device joins an IoT environment, an authenticated IoT device must verify whether or not the new device is within the boundary of operations using their site following the steps below:

1) An authenticated IoT device in the system scan for the beacon frame of every Wi-Fi access point in the environment then collect it. Also, it measure the RSSI value of every presented access point as follows:

$$B_1 \begin{matrix} \bullet \\ Tuple_1 = \{SSID_1, BSSID_1, RSSI_1\} \\ Tuple_2 = \{SSID_2, BSSID_2, RSSI_1\} \\ \bullet \\ \bullet \\ \bullet \\ Tuple_n = \{SSID_n, BSSID_n, RSSI_n\} \end{matrix} \quad (1)$$

2) When a new device wants to be added to the network the authenticated device verifies whether the new device is within the boundary of operations.
3) Similar to step, 1 the new device scan for the beacon frame of every Wi-Fi access point in the environment and then collect it. Also, it measures

the RSSI value of every presented access point as follows:

$$Tuple_1 = \{SSID_1, BSSID_1, RSSI_1\}$$
$$B_2 \quad Tuple_2 = \{SSID_2, BSSID_2, RSSI_1\}$$
$$\vdots$$
$$Tuple_n = \{SSID_n, BSSID_n, RSSI_n\} \quad (2)$$

Then it sends it to the authenticated device.

4) When the authenticated device receives the data in step 3, it calculates the Euclidean distance between the two data sets, B1 and B2, using equation 3:

$$D = \sqrt{\sum_{i=1}^{n} [B_1 Tuple_i(RSSI) - B_2 Tuple_i(RSSI)]^2} \quad (3)$$

Where $D$ is the Euclidean distance.

If $D$ is below a certain defined threshold, the proximity authentication is successful. If $D$ is above this threshold, proximity authentication fails. The threshold is determined based on a calibration experiment. Such calibration can be done by the vendor for similar devices. Alternatively, it can be calculated at the area of operation for different devices.

The proposed system facilitates proximity authentication for multiple nodes. Authenticated devices are able to verify new device proximity before authenticating them to the network. This is done using the threshold, which is determined by the Euclidean distance in equation 3. When the calculated distance is below the threshold, it means that the accepted proximity reflects the most efficient data transmission in the ad hoc network. Also, it enforces security in the system to make sure far or imposter devices are not able to authenticate. Figure 1 shows how authentication works in the proposed system.

As can be seen in Figure1 (a). Node 1 is an authenticated node in the system. The color green denotes an authenticated node. Node2 enters the system and wants to be authenticated. It scans for Wi-Fi Beacon and broadcasts this data in the request to join the network. Node1 received this request along with the scan data. Node1 verifies whether Node2 meets the proximity thresholdby calculating the Euclidean distance and comparing the distance with the pre-defined threshold. Figure1(b) shows that the authentication is successful. In Figure1 (c), aNode 3 enters the area. Node3 is more proximate to Node2 than Node1. Node 3 broadcasts the request to join the network along with the Wi-Fi scan data, which is received by Node1 and Node2. Both Node1 and Node 2 verify whether Node3 meets the proximity threshold in reference to these nodes. Figure1(d) shows that it met the threshold in reference to Node2 but not Node1. For this case, the new device meat the threshold with two devices. It connects to the device with the least Euclidean distance.

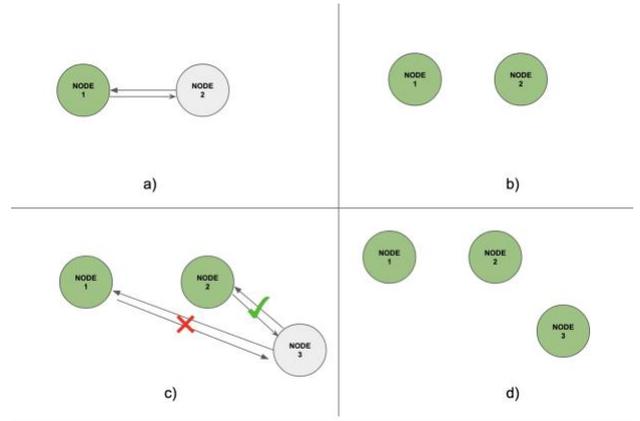

Fig. 1. Authenticating new IoT nodes

## VI. EXPERIMENTS

To test the proposed system, two experiments were conducted. Experiment 1 calculates the accuracy of the proposed proximity-based authentication. Experiment 2 simulates the system operation with several nodes.

### A. Two-device proximity authentication

In this experiment, two Raspberry Pis were placed within two meters of each other. Each Raspberry Pi scanned and collected the Wi-Fi beacon frames and RSSI in the area following equation 1 and equation 2. The collected data was used to calculate the value of the threshold using equation 3. In the experiment, the ten access points with the highest RSSIs were used in the equation. To test the proposed system, the two Raspberry Pis were placed at ten different locations around the building. Ten authentications were attempted at each location. Five of these attempts were conducted when the two devices were less than two meters apart. Then, five more times where the two devices are more than two meters apart. The data was collected for each attempt and the Euclidean distance was calculated using equation3. This distance was compared to the threshold to determine successful and failing authentications. The results of the experiment are shown in Table I.

TABLE I
SUCCESS RATE

| N = 100 | Actual | |
|---------|--------|--------|
|         | Success | Failure |
| True    | TS = 45.54% | TF = 42.36% |
| False   | FS = 4.46% | FF = 7.64% |

The experiment returned authentication accuracy of 87.9%. The accuracy can be increased if the tolerance of the threshold value is increased. When 20% tolerance is considered, the accuracy jumps to 94.5%.

### B. Several nodes simulation

This experiment is to simulate the system's operations when several nodes are involved. The simulation was conducted utilizing python. The threshold calculated in the experiment above was considered. A function was written to perform the authentication following the steps above. The nodes in the simulation were configured with these Actual RSSI values at ten different locations collected in the experiment above. The simulation had each node to authenticate with the other through an iteration. Each node in the simulation would attempt to connect to the node where there are the nodes that meet the threshold and with the least Euclidean distance. In the simulation, each device is connected to a node with the least distance, as shown in Figure 2. The experiment returned authentication accuracy of 90%.

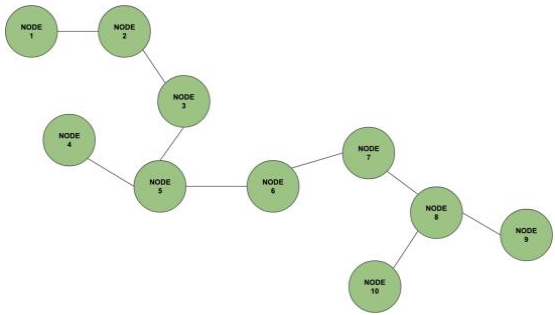

Fig. 2. Authenticating new IoT nodes

## VII. SYSTEM ANALYSIS

### A. Continuous Authentication

Continuous authentication is a feature where the authentication process every predefined time. The service provider will check continuously if the two communicated IoT devices are in the same IoT network in order to maintain the section. The frequency with which the IoT devices collect and send the new data (SSID, BSSID, RSSI values) can be varied based on requirements and/or an administration's desire. The continuous authentication will check if both communicated IoT devices are still in the same IoT environment in order to keep the session alive. If not, the session can be automatically terminated. This feature could mitigate and, at some level, prevent some types of cyber-attacks (more details are provided in the next section).

### B. Usability

Due to the proposed scheme using the existing IEEE 802.11 infrastructure (i.e., Wi-Fi access points already in the IoT environment, network interface cards already in IoT devices), no new hardware is required to be installed. Moreover, the proposed scheme is considered to be readily applicable in a scalable manner because it relies on ubiquitous Wi-Fi access points. The internet can be found almost everywhere people live [18]. Due to its ubiquitous nature, the internet is an essential and robust platform for education, business, and entertainment. It has been noted that locations with reliable internet connectivity are also where access points are commonly established [19].

## VIII. SECURITY ANALYSIS

### A. Environment Simulation Attack

A chance of simulating the IoT Environment is possible. The attacker scans the IoT environment (i.e., beacon frame characteristics and RSSI value) and then replicates it elsewhere, where the attacker has full control of the simulated environment. In the proposed research, the IoT environment can be scanned and replicated; however, the attacker's IoT device's unique identifier (e.g., IMEI, UUID, MAC address, etc.) will not match the IoT's unique identifier in the database. Moreover, the admin- istration will be notified because the authentication entity will reject the request. Also, The continuous authentica- tion feature will mitigate and, at some level, will prevent the attack.

### B. Insider attacks

An attacker can be inside the IoT environment, which means he/she will be able to scan the IoT environment (i.e., beacon frame characteristics and RSSI value) and then utilizes it. Beginning inside the IoT environment and scanning it, is something easy to obtain. However, the continuous authentication feature will mitigate and, at some level, will prevent the attack. In addition, the service provider will notice that the malicious IoT device's unique identifier does not match the one in its database, which results in rejecting the request and notifying the administrator.

## IX. CONCLUSION

IoT is one of many buzzwords in Information Technology (IT), which will transform our daily lives into smart systems. To guarantee the security of the wireless communications in an IoT environment, devices must build trust in the identity of each other, *authentication*. This paper proposes a technique to authenticate IoT devices in ad hoc networks to verify proximity. This is done in a way that only devices within a certain distance from other authenticated IoT devices will be able to connect to the network. Meanwhile, devices that are far from an authenticated device or not physically in the area will fail in the proximity authentication. The proposed system enforces security in ad hoc IoT networks. Also, it figures the more suitable device to connect to in an ad hoc network that would reflect the most suitable Radio frequency conditions to communicate. The experiment showed an adequate accuracy of proximity authentication that can be increased with configuring the tolerance in the threshold.